\documentclass[sigconf,natbib=true]{acmart}
\usepackage{subfigure}
\usepackage{soul}
\usepackage{multirow}
\theoremstyle{plain}

\usepackage{makecell}
\makeatletter
\newif\if@restonecol
\makeatother

\usepackage[ruled,vlined,linesnumbered]{algorithm2e}
\let\oldnl\nl
\newcommand{\nonl}{\renewcommand{\nl}{\let\nl\oldnl}}
\AtBeginDocument{%
  \providecommand\BibTeX{{%
    \normalfont B\kern-0.5em{\scshape i\kern-0.25em b}\kern-0.8em\TeX}}}

\setcopyright{acmcopyright}
\copyrightyear{2018}
\acmYear{2018}
\acmDOI{XXXXXXX.XXXXXXX}

\acmConference[Conference acronym 'XX]{Make sure to enter the correct
  conference title from your rights confirmation emai}{June 03--05,
  2018}{Woodstock, NY}
\acmBooktitle{Woodstock '18: ACM Symposium on Neural Gaze Detection,
 June 03--05, 2018, Woodstock, NY} 
\acmPrice{15.00}
\acmISBN{978-1-4503-XXXX-X/18/06}

\title{On the Adaptation to Concept Drift for CTR Prediction}

\author{Congcong Liu*, Yuejiang Li*, Fei Teng,Xiwei Zhao, Zhangang Lin,Jinghe Hu,Jingping Shao }
\affiliation{%
  \institution{JD.com}
  \country{Beijing, China}
  }
  
\email{ {liucongcong25,liyuejiang1,tengfei49,zhaoxiwei,linzhangang,hujinghe,shaojingping}@jd.com }

\begin{document}

\begin{abstract}
Click-through rate (CTR) prediction is a crucial task in web search, recommender systems, and online advertisement displaying.
In practical application, CTR models often serve with high-speed user-generated data streams, whose underlying distribution rapidly changing over time.
The concept drift problem inevitably exists in those streaming data, which can lead to performance degradation due to the timeliness issue.  
To ensure model freshness, incremental learning has been widely adopted in real-world production systems.
However, it is hard for the incremental update to achieve the balance of the CTR models between the adaptability to capture the fast-changing trends and generalization ability to retain common knowledge.
In this paper, we propose adaptive mixture of experts (AdaMoE), a new framework to alleviate the concept drift problem by statistical weighting policy in the data stream of CTR prediction.
The extensive offline experiments on both benchmark and a real-world industrial dataset, as well as an online A/B testing show that our AdaMoE significantly outperforms all incremental learning frameworks considered.
\end{abstract}

\maketitle





\section{Introduction}

Click-through rate (CTR) prediction has been widely explored in online advertising and recommender systems with deep learning models, mainly focusing on feature interaction\cite{wang2017deep,cheng2016wide,guo2017deepfm,qu2016product} and user behavior modeling \cite{zhou2018deep,zhou2019deep,xiao2020deep,huang2021deep}.
These deep CTR models are usually trained with offline batch mode.

\begin{figure*}
  \centering
  \subfigure[6-hour vs. 6-hour similarity]{
  \includegraphics[width=0.4\textwidth]{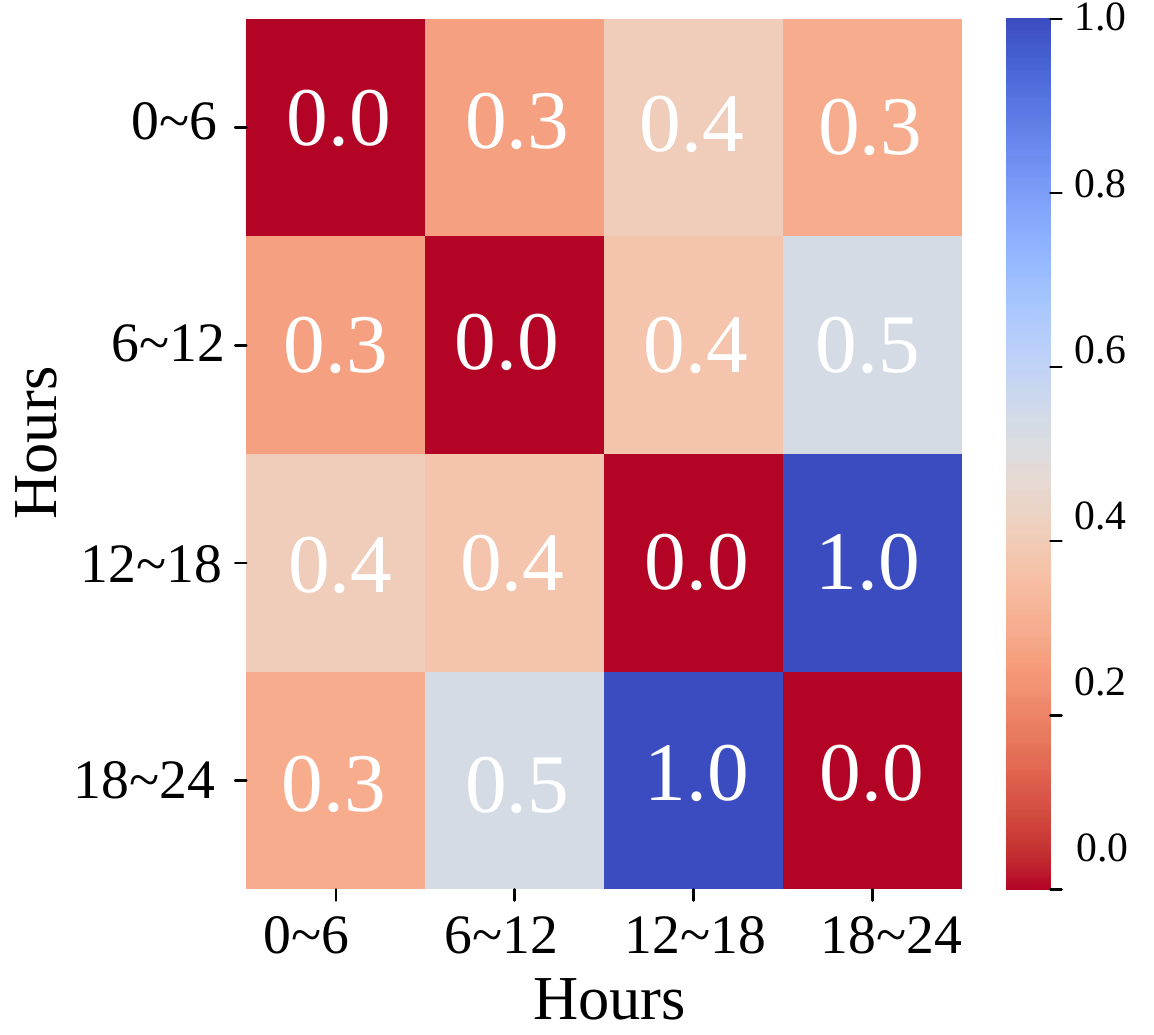}
  }
  \subfigure[Variation of CTR]{
  \includegraphics[width=0.38\textwidth]{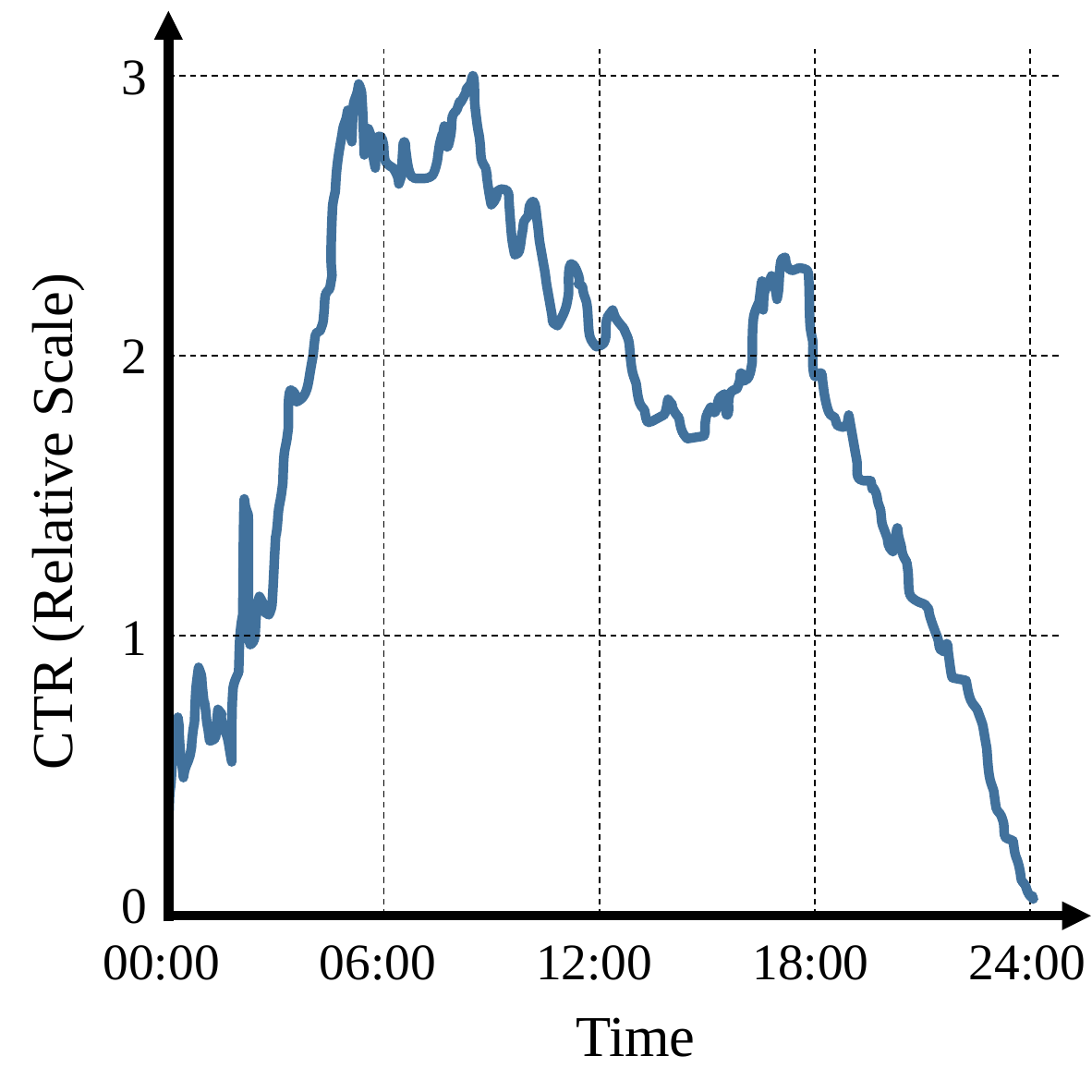}
  }
  \caption{Empirical observation of concept drift. (a) Impressed ad samples are grouped every six hours, and the energy distance \cite{sugiyama2015introduction} between groups of samples is calculated and unified.
  Larger value denotes larger dissimilarity.
  (b) CTR in real production within a day is shown in relative scale.}
  \label{fig:drift}
\end{figure*}

However, in real-world production systems such as online advertising platforms, CTR models often serve with high-speed streaming data generated from a huge amount of users.
One common challenge is to deal with rapid changes in the distribution of the data stream over time, referred to as the concept drift problem \cite{tsymbal2004problem,lu2018learning}. Figure. \ref{fig:drift} provides an empirical observation of concept drift with hours-hours similarity and temporal variation of CTR values in a real-world online advertising system.

To guarantee model freshness, incremental learning has been applied to real-world production systems \cite{wang2020practical}. 
However, vanilla incremental learning methods are not able to address the \textit{stability-plasticity dilemma} \cite{gepperth2016incremental}.
Slow updates achieve stability but reduce model reactivity.
Fast adaptation with large steps can help the model keep up with concept drift, but decreases its robustness.

Many efforts have been devoted to learning and mining of concept drifting stream data \cite{Widmer1996learning,kolter2007dynamic,bifet2007learning,elwell2011incremental,yang2017dynamic}.
However, these models require an additional buffer to store historical data or need extra drift detection module to work properly.
The memory inefficiency and computational complexity of these models can be problematic in the application of online CTR inference \cite{yang2019adaptive}.

Without extra cache and explicit detection for drifting data, many recent works investigate a more adaptive way to update models according to new patterns \cite{wang2020practical,yang2019adaptive,street2001streaming}.
The work in \cite{wang2020practical} proposes IncCTR which applies incremental learning to CTR prediction task.
IncCTR employs knowledge distillation to balance the learned knowledge from the previous model and that from incoming data.
However, IncCTR uses a single and fixed-structured model, which limits the adaptation capacity to streaming data with potential concept drift.
Ramanath \textit{et al.} introduces Lambda Learner \cite{ramanath2021lambda}, an incremental learning framework of generalized additive mixed effect models. 
However, its update rule is tailored for simple linear models with less pressure on memory and computational resource requirements.
A more popular way is to leverage ensemble learning paradigm \cite{gomes2017survey,krawczyk2017ensemble,yang2019adaptive}.
The work in \cite{yang2019adaptive} proposes incremental adaptive deep model (IADM) which
starts from a shallow network, evolves to a deep network, and ensemble the outputs from different depth for streaming data.
However, when the network evolves sufficiently deep, it is still hard to converge to optimum when training with streaming data.

Mixture of experts (MoE) \cite{jacobs1991adaptive} is a canonical deep ensemble learning structure. 
In the MoE model, all expert networks share the same backbone network which extracts the representations of input, and a gate network is designed to decide the aggregation weight of the output of each expert.
Although the MoE model yields significant performance in various applications, training MoE can be challenging with possible dead gate issues \cite{ShazeerMMDLHD17,eigen2013learning}.

In this paper, we propose a novel framework of incremental learning, AdaMoE, which decouples the update of aggregation module from regular MoE network training scheme for CTR prediction.
To update the aggregation module for fast adaptation, we theoretically derive
the optimal updating policy upon the performance measurements of the prediction from experts, rather than inefficient back propagation via iterative gradient descent.
This novel statistical update operation with closed-form solution greatly eases the convergence difficulty for MoE based methods, since knowing part of the parameters apparently makes the estimation of remaining parameters easier \cite{makkuva2019breaking}.

To validate the effectiveness and efficiency of the proposed method in concept drifting data streams, we conduct extensive experiments against various competing baselines for CTR prediction on real-world production dataset and a chronologically reorganized benchmark.
We further demonstrate the performance of our AdaMoE through a rigorous online A/B test in an online advertising system.
An intuitive qualitative analysis of CTR values over time is provided to illustrate the superiority of AdaMoE in handling concept drift.

\begin{figure*}
 \centering
  \includegraphics[width=0.6\textwidth]{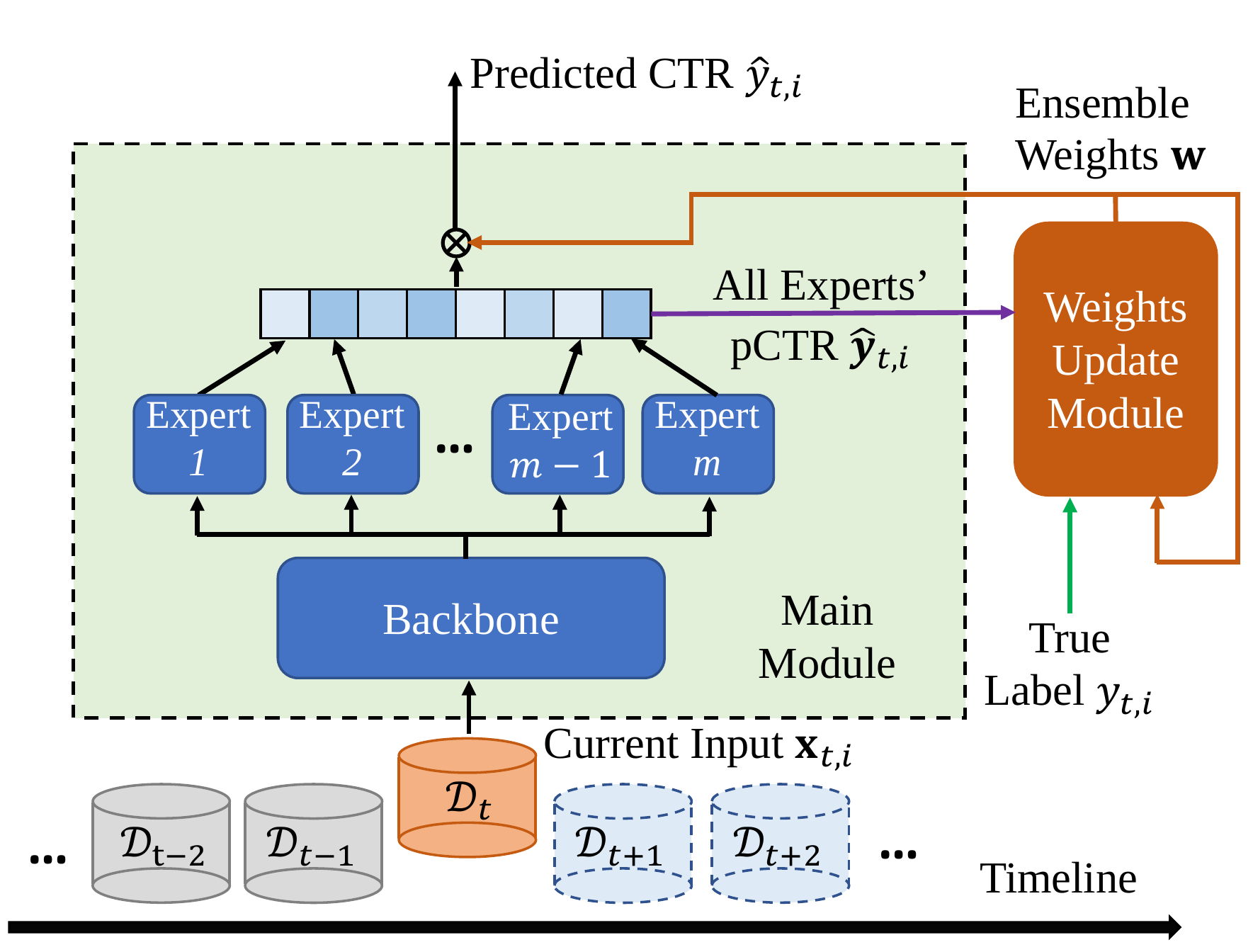}
  \caption{An overview of AdaMoE.}
  \label{fig:module}
\end{figure*}



The contributions of this paper is summarized below:
 \begin{itemize}
     \item We proposed an innovative incremental learning framework, AdaMoE, for serving high-velocity user-generated streaming data. 
     \item Theoretical derivation has been conducted for incremental update of CTR models in concept drifting streaming data. Approximation capability and generality analysis of the proposed method are provided.  
     \item We reorganized a widely used benchmark chronologically to fill the absence of the public dataset studying concept drift problem in CTR prediction. Both reorganized dataset and source code \footnote{https://github.com/Yuejiang-li/FuxiCTR/tree/liyuejiang/develope} has been released to the community.  
     \item We achieve significant improvements on both public benchmark and a real-world industrial dataset over all incremental learning baselines. A rigorous A/B test further demonstrates the excellent performance over a highly optimized baseline models. The proposed AdaMoE had been deployed in a real-world advertising system, serving hundreds of millions of active users.
 \end{itemize}


\section{Methodology}

\subsection{Problem Definition}
In this work we consider CTR prediction with stream data.
Let $\{\mathcal{D}_t\}_{t=1}^{+\infty}$ be the stream data, where $\mathcal{D}_t = \left\{\mathbf{x}_{t, i}, y_{t, i}\right\}_{i=1}^{N_t}$ denotes the training data at time step $t$ and $(\mathbf{x}_{t, i}, y_{t, i})$ are generated i.i.d. from the density $P_t(X, y)$.
Here, $\mathbf{x}_{t, i}\in \mathbb{R}^f$ represents the data features of the $i$-th impressed advertisement at time step $t$, and
$y_{t, i} \in \{0, 1\}$ is the ground truth label indicating whether the user clicks the item.
\subsection{Influence of Concept Drift}
Concept drift problem refers to the phenomenon in which the statistical properties of data features and targets change in unforeseen ways over time\cite{lu2018learning}, and it happens at time step $t$ if $P_t(X, y)\neq P_{t+1}(X, y)$ \cite{ramirez2017survey}. 
In the context of CTR prediction, the distribution of user-generated data steams consistently changed over time due to drifting user interests or user population.

\textbf{The influence of concept drift on model performance.} 
Suppose there is a predictor $F_t$ that predict the label for $\forall \mathbf{x}_t\sim P_t(\mathbf{x})$ perfectly, and the best predictor under some restrict class is $F_t^*$. However, under various restrictions, such as complexity, non-convexity and model selection, we can only find $\tilde{F}_t^*$ in real practice, which is an approximation of $F_t^*$. We can decompose the error of $\tilde{F}_t^*$ as


\begin{equation}
    \epsilon(\tilde{F}_t^*) - \epsilon(F_t) = \underbrace{\epsilon(\tilde{F}_t^*) - \epsilon(F_t^*)}_{\text{estimation error}} + \underbrace{\epsilon(F_t^*) - \epsilon(F_t)}_{\text{approximation error}} \label{eqn:error},
\end{equation}
where the estimation and approximation error reflects the inability of the learning algorithm and the lack of finite data, model selection and so on respectively. If concept drift occurs, the approximation error becomes $\epsilon(F_t^*) - \epsilon(F_{t+1})$, which is larger in most cases, thus the performance of the model is reduced. Therefore, when the metrics of the model such as AUC drops significantly, we should pay attention to whether there occurs concept drift.

\textbf{Why MoE structure can alleviate concept drift? }
The vanilla incremental learning methods (single estimator, i.e., one-expert-model) lacks robustness to changes over time. The basic idea is to construct $m$ experts to enhance robustness and accelerate adaptation to the new distribution.
The following inequality holds\cite{zeevi1998moe}:

\begin{equation}
    ||F-F^{(m)*}||_p\leq cm^{-\frac{r}{d}}, 1 \leq p < \infty,
\end{equation}
where $d$ is the input dimension, $F$ is a Sobolev class function of order $r$ in $L_p$ norm and $F^{(m)*}$ is the MoE function of $m$ experts. This inequality gives an upper bound of the approximation error, indicates the convergence of MoE, and
gives an approximation that $F^{(m)*}$ can uniformly approximate $F$ at a rate of at least $\mathcal{O}(cm^{-\frac{r}{d}})$. Thus, if $F_t$ changes to $F_{t+1}$, the MoE model $F^{(m)*}_t$ can adapt to the new distribution  therefore alleviate concept drift.


\subsection{Framework}

In this work, we propose the adaptive mixture of experts model (AdaMoE) in CTR prediction with stream data. As illustrated in Fig. \ref{fig:module}, the proposed AdaMoE consists of two modules: the \textit{main module} that uses the input data feature to predict the CTR, and the \textit{weights update module} that dynamically computes and updates the aggregation weights of all experts' predictions.

\subsubsection{Main Module}
In the main module, the representation $\mathbf{e}_{t, i}$ of sample $\mathbf{x}_{t, i}$ is first extracted with a backbone model, which can be DCN \cite{wang2017deep} or DIN \cite{zhou2018deep}, etc., and fed into $m$ experts.
Each expert is an MLP network with a sigmoid activation.
All experts convert $\mathbf{e}_{t, i}$ to their own predicted CTR (pCTR) $\hat{y}_{t, i}^{(k)}$.
Assume that all samples use the same aggregation weights, $\mathbf{w}_t = [w^{(1)}_{t}, \cdots, w^{(m)}_{t}]^T \in [0,1]^{m}$, to aggregate the output of experts, $||\mathbf{w}||_1=1$. Denote $\hat{\mathbf{y}}_{t, i} = [\hat{y}_{t, i}^{(1)}, \cdots \hat{y}_{t, i}^{(m)}]^T$, the pCTR is given by
\begin{equation}
  \hat{y}_{t, i} = \hat{\mathbf{y}}_{t, i} \cdot \mathbf{w}_{t} = \sum_{k=1}^{m} w^{(k)}_{t} \hat{y}_{t, i}^{(k)}.\label{eqn:pctr}
\end{equation}

In AdaMoE, we separate the training of the main module and the compute of experts' aggregation weights.
To train the main module, for the sample $\mathbf{x}_{t, i}$, we calculate the cross entropy loss
\begin{equation}
  \mathcal{L}_{CE}(t, i) = -(y_{t, i} \cdot \log(\hat{y}_{t, i}) + (1 - y_{t, i}) \cdot \log(1 - \hat{y}_{t, i})).\label{eqn:celoss}
\end{equation}
The main module is optimized with gradient descent, but the data at time $t$ is used only once as the data is in stream form.


\subsubsection{Weights Update Module}
\label{sec:weight-update}
\begin{algorithm}[t]
\SetAlgoLined
\KwIn{Data $\mathcal{D}_t = [\mathbf{x}_t, \mathbf{y}_t] = \left\{\mathbf{x}_{t, i}, y_{t, i}\right\}_{i=1}^{N_t}$.}
\KwOut{Predicted CTR $\hat{y}_{t, i}$ for each impressed advertisement.}

  {\nonl\footnotesize\texttt{// Forward: predict CTR}}\;

  $\hat{\mathbf{y}}_{t, i} \leftarrow \text{MainModule}(\mathbf{x}_{t, i})$; \hspace{3em}{\footnotesize\texttt{// Collect pCTR of experts}}\;

  $\hat{y}_{t, i} \leftarrow \hat{\mathbf{y}}_{t, i} \cdot \text{stopgrad}(\mathbf{w}_{t-1})$;\hspace{3em}{\footnotesize\texttt{// Aggregate pCTR of experts}}\;

  {\nonl\footnotesize\texttt{// Backward: update main module and aggregation weights}}\;

  $\tilde{\mathbf{y}}_{t, i} \leftarrow y_{t, i}\cdot \hat{\mathbf{y}}_{t, i} + (1-y_{t, i})\cdot (1-\hat{\mathbf{y}}_{t, i})$; 
  
  $\mathbf{w} \leftarrow \frac{1}{N_t}\sum_{i=1}^{N_t}\frac{\tilde{\mathbf{y}}_{t, i}}{||\tilde{\mathbf{y}}_{t, i}||_1}$; \hspace{3em}{\footnotesize\texttt{// Compute aggregation weights}}\;

  $\mathbf{w}_{t} \leftarrow \lambda\mathbf{w_{t - 1}} + (1-\lambda)\mathbf{w}$; \hspace{3em}{\footnotesize\texttt{// Update aggregation weights}}\;

  \caption{AdaMoE for CTR Prediction.}
\end{algorithm}

In this work, we aim to efficiently compute and update the aggregation weights of experts' predictions.
To achieve this, the aggregation weights should minimize the loss function and are easily computed.
We consider the computation of the aggregation weight at time step $t$ and omit script $t$ for simplicity. Based on Eqn. (\ref{eqn:pctr})(\ref{eqn:celoss}), the total loss to minimize at time $t$ is

\begin{equation}
    \mathcal{L} = -\sum_{i=1}^{N}(y_i\cdot\log\sum_{k=1}^{m} w^{(k)} \hat{y}_{i}^{(k)} + (1-y_i)\cdot\log\sum_{k=1}^{m} w^{(k)} (1-\hat{y}_{i}^{(k)})).\label{eqn:loss}
\end{equation}
Let $\tilde{y}_i^{(k)} = y_i\cdot \hat{y}_{i}^{(k)} + (1-y_i)\cdot (1-\hat{y}_{i}^{(k)})$, $\tilde{\mathbf{y}}_{i} = [\tilde{y}_{i}^{(1)}, \cdots \tilde{y}_{i}^{(m)}]^T$, 
\begin{equation}
    \mathcal{L} = -\sum_{i=1}^{N}(\log\sum_{k=1}^{m} w^{(k)} \tilde{y}_{i}^{(k)}) = -\sum_{i=1}^{N}(\log(\tilde{\mathbf{y}}_{i} \cdot \mathbf{w})).\label{eqn:loss_all}
\end{equation}
The target is to minimize $\mathcal{L}$.
For simplicity, we ignore the partial derivative $\frac{\partial w^{(j)}}{\partial w^{(i)}}$ for $\forall i \neq j $. In fact, $\frac{\partial w^{(j)}}{\partial w^{(i)}}$ tends to $0$ as \#exp. increases. Then, $\mathcal{L}' := \frac{\partial \mathcal{L}}{\partial w} = -\sum_{i=1}^{N}\frac{\tilde{\mathbf{y}}_{i}}{\tilde{\mathbf{y}}_{i} \cdot \mathbf{w}} \leq 0$ since $\tilde{\mathbf{y}}_{i}$ and $\mathbf{w}$ are not less than $0$. Thus, we try to find $\mathbf{w}$ that can make $\mathcal{L}'$ closest to $0$ and maximize $\mathcal{L}'$ as much as possible.
Noticed that $||\mathbf{w}||_2 \leq ||\mathbf{w}||_1 =1$ by Cauchy-Schwarz inequality, we can write $\mathcal{L}'$ as 
\begin{equation}
  \mathcal{L}' = -\sum_{i=1}^{N}\frac{\tilde{\mathbf{y}}_{i}}{||\tilde{\mathbf{y}}_{i}||_2||\mathbf{w}||_2\cos{\theta_i}} \leq -\sum_{i=1}^{N}\frac{\tilde{\mathbf{y}}_{i}}{||\tilde{\mathbf{y}}_{i}||_2}\cdot\frac{1}{\cos{\theta_i}}, \label{eqn:simple}
\end{equation}
where $\theta_i$ is the angle between $\tilde{\mathbf{y}}_{i}$ and $\mathbf{w}$.
Instead of directly finding $\mathbf{w}$ that maximize $\mathcal{L}'$, we maximize the right hand side of eqn.  (\ref{eqn:simple}). For $i$-th impressed advertisement, $\frac{1}{\cos{\theta_i}}$ reachs its minimum when $\cos{\theta_i} = 1$, i.e., $\tilde{\mathbf{y}}_{i}$ and $\mathbf{w}$ are collinear and $\mathbf{w} = \frac{\tilde{\mathbf{y}}_{i}}{||\tilde{\mathbf{y}}_{i}||_1}$. To be fair to each impressed advertisement, let 

\begin{equation}
    \mathbf{w} = \frac{1}{N}\sum_{i=1}^{N}\frac{\tilde{\mathbf{y}}_{i}}{||\tilde{\mathbf{y}}_{i}||_1}, \label{eqn:w}
\end{equation}
$\mathbf{w}$ is initialized as $\mathbf{w}_0=[\frac{1}{m}, \frac{1}{m},...,\frac{1}{m}]^T \in \mathbb{R}^{m}$. Monte Carlo simulation is conducted to examine approximation capability. $\mathcal{L}$ computed with eqn. \ref{eqn:w} is approximately equals to the minimum of $\mathcal{L}$ at 90\% significance level. 


\textbf{Weight Update Rule. }\label{sec:weight-update}
The aggregation weight (eqn. (\ref{eqn:w})) considers only the information in time step $t$. Considering historical information, we introduce a decay factor $\lambda$ into the update rule of aggregation weight, and the aggregation weights at time step $t$ is 
\begin{equation}
    \mathbf{w_t} = \lambda\mathbf{w_{t - 1}} + (1 - \lambda)\mathbf{w},\text{where }\mathbf{w} = \frac{1}{N_t}\sum_{i=1}^{N_t}\frac{\tilde{\mathbf{y}}_{t, i}}{||\tilde{\mathbf{y}}_{t, i}||_1},
\end{equation}
where decayed factor $\lambda \in [0, 1)$. $\lambda = 1$ is not considered because we ignore the case where $w$ is not updated with current information.
To sum up, 
we summarize the process of AdaMoE in Algorithm 1. Besides, the proposed approach is easy to be generalized to multi-class setting through the similar way of defining $\mathcal{L}_{CE}(i)$ and $\tilde{y}_{i}^{(k)}$.

\textbf{Why AdaMoE works?} The $k$-th expert works well at time step $t$ when its prediction $\hat{y}_{t, i}^{(k)}$ is close to the ground truth $y_{t, i}$, i.e. $\tilde{y}_{t, i}^{(k)}$ is close to $1$. AdaMoE analytically gives the aggregation weights that minimize the loss function, and gives larger weights to those experts that work better
. Thus, AdaMoE can find the better-performing expert from $m$ experts and give larger weights fast and precisely. Besides, AdaMoE also considers the history information, which contains the performance of each expert at historical time steps.

\section{Experiments}




\subsection{Datasets and Evaluation Metrics}
\textbf{Industrial Dataset}.
The industrial dataset is extracted from the user logs of one of world's largest E-commerce companies.
The extracted dataset includes about 4.5 billion ad impression records within one day.
The collected data is sorted in chronological order and divided into datasets.
Each dataset $\mathcal{D}_t$ contains 2048 samples.

\textbf{Public Dataset}. Due to the absence of the public dataset to study concept drift in CTR Prediction task, here we adopt the Avazu dataset\footnote{\url{https://www.kaggle.com/c/avazu-ctr-prediction/data}}, which includes the timestamp information of impressed advertisement. There are 40,428,967 records over 10 days.
We use the records of the first 3 days to pretrain models. For the records of the rest 7 days, we sort them in chronological order, and split them into stream dataset $\mathcal{D}_t$ by a fixed interval of an hour due to the timestamp granularity in the Avazu dataset.
In the experiments, the stream dataset $\mathcal{D}_t$ is consumed sequentially. Stream dataset $\mathcal{D}_t$ is first evaluated by the model trained with previous data. Then, $\mathcal{D}_t$ is used to train the model. Our code, including data preprocessing and model implementation, will be publicized on acceptance.

\begin{table}[t]
  \caption{Overall AUC on Industrial and Avazu dataset over 3-Run, std$\approx10^{-4}$. \#(Exp.) denotes number of experts.}
  \label{tab:main-res}
  \centering
  \small
\resizebox{1\columnwidth}{!}{ 
\begin{tabular}{c|c|cccc}
   \toprule
   Dataset & \#(Exp.) & IncCTR & MoE & IADM & AdaMoE \\
   \midrule
   \multirow{4}{*}{Industrial(DCN)} & 3 & 0.7569 & 0.7578 & 0.7582 & \textbf{0.7590} \\
                        & 6 & 0.7571 & 0.7580 & 0.7578 & \textbf{0.7597} \\
                        & 9 & 0.7568 & 0.7579 & 0.7576 & \textbf{0.7593} \\
                        & 12 & 0.7563 & 0.7573 & 0.7571 & \textbf{0.7585} \\
   \midrule
   \multirow{4}{*}{Industrial(DIN)} & 3 & 0.7631 & 0.7635 & 0.7643 & \textbf{0.7663} \\
                        & 6 & 0.7635 & 0.7638 & 0.7642 & \textbf{0.7664} \\
                        & 9 & 0.7633 & 0.7636 & 0.7640 & \textbf{0.7660} \\
                        & 12 & 0.7628 & 0.7631 & 0.7633 & \textbf{0.7657} \\
    \midrule
    \multirow{4}{*}{Avazu} & 3 & 0.7601 & 0.7599 & 0.7614 & \textbf{0.7618} \\
                        & 6 & 0.7601 & 0.7603 & 0.7597 & \textbf{0.7620} \\
                        & 9 & 0.7590 & 0.7602 & 0.7157 & \textbf{0.7624} \\
                        & 12 & 0.7575 & 0.7598 & 0.7170 & \textbf{0.7619} \\
   \bottomrule
\end{tabular}}
\end{table}
\begin{figure*}[t]
  \centering
  \subfigure[AUC/10Min of different models]{
  \includegraphics[width=0.45\textwidth]{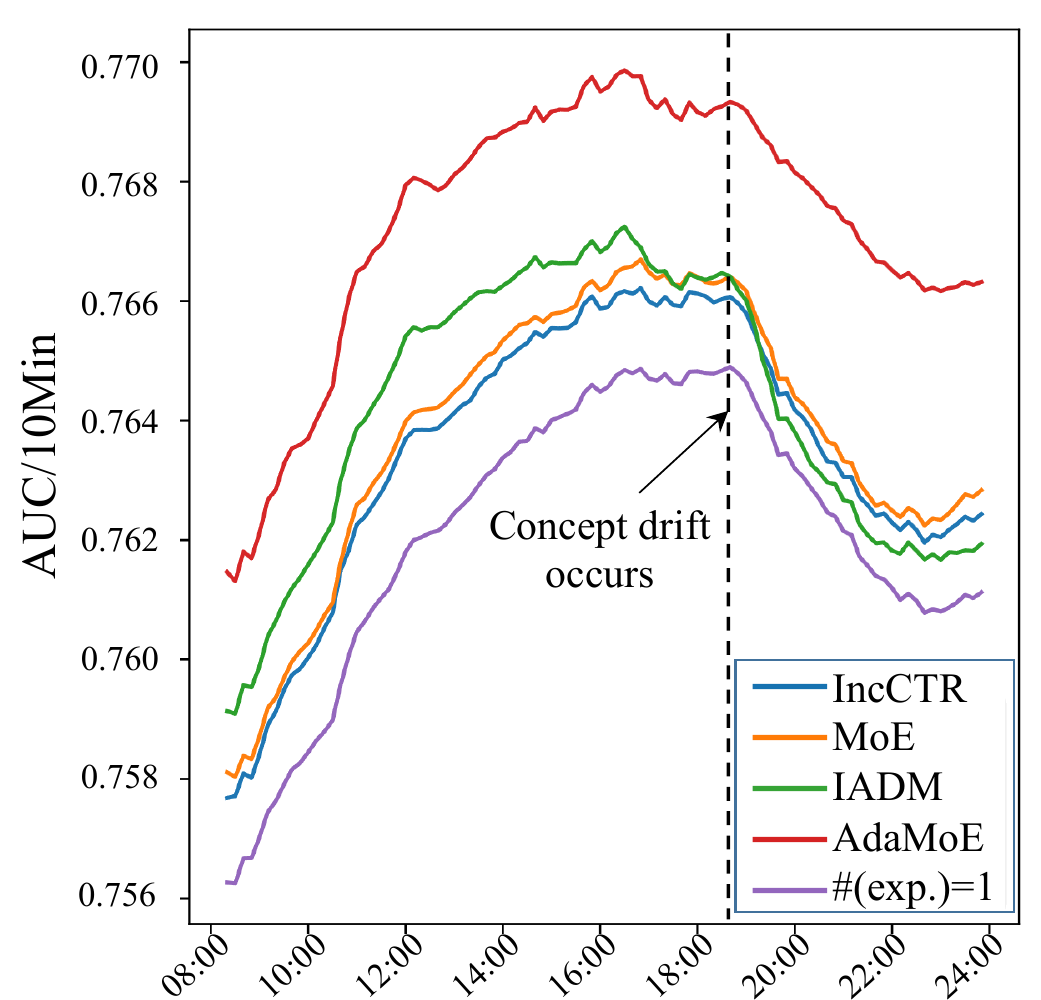}
  \label{fig:auc-models}
  }\hspace{-10pt}
  \subfigure[AUC/10Min of AdaMoE with different $\lambda$]{
  \includegraphics[width=0.45\textwidth]{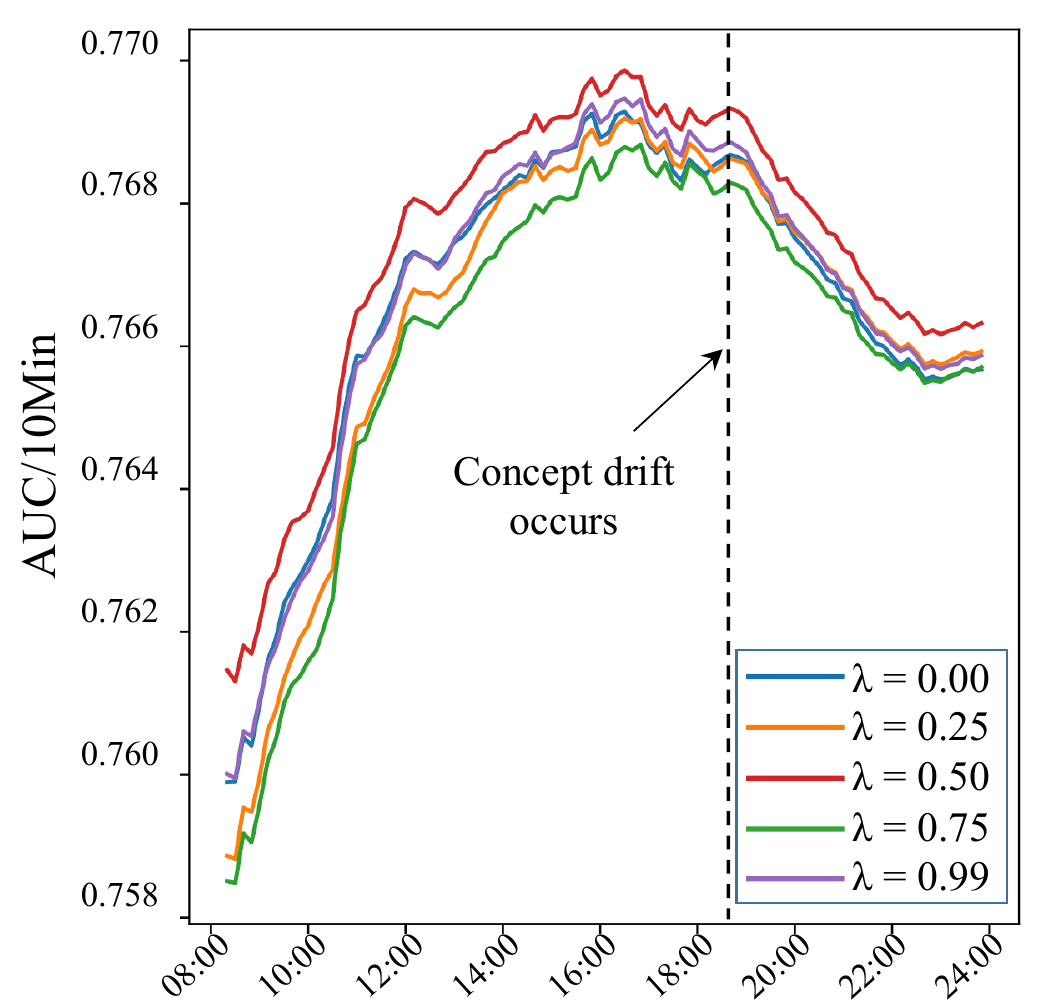}
  \label{fig:auc-lambda}
  }
  \caption{AUC/10Min of (a) different models and (b)  $\lambda$.}
  \label{fig:auc-change-results}
\end{figure*}

\textbf{Evaluation Metrics}.
We adopt AUC as the evaluation metric.
In our setting, data is in the stream form and fed for training only once.
At each time step $t$, we use the first 80\% of data as training samples $\mathcal{D}_t^{train}$, and the rest 20\% as testing samples $\mathcal{D}_t^{test}$. The model is first trained with $\mathcal{D}_t^{train}$
then applied to $\mathcal{D}_t^{test}$.
The output pCTRs and the ground truth click labels of samples in $\mathcal{D}_t^{test}$ are stored to
calculate the \textit{overall AUC}.
Furthermore, to better investigate the performance of models with stream data, we collect the stored pCTR and label of test samples every 10 minutes to calculate \textit{AUC/10Min}.

\subsection{Baseline and Settings}

We compare AdaMoE with the following baseline models.

$\bullet$ \textit{IncCTR}: 
the pCTR is simply the average of all experts' output. 

$\bullet$ \textit{MoE}: 
the outputs of experts are aggregated by a gate network. The gate network used for the MoE baseline is a two-layer MLP with 64 and 32 neurons, respectively.

$\bullet$ \textit{IADM}: which stack ``experts'' in the depth direction, and the output of experts are also aggregated by a gate network. 

For industrial dataset, we use DCN \cite{wang2017deep} and DIN \cite{zhou2018deep} as the backbone, respectively. For public dataset, we use DCN as backbone due to the absence of user behavior data in avazu dataset.
The output dimension of the backbone is 1024.
Each expert is a 2 layer MLP with [512, 256] hidden units and ReLU activation, combining a linear layer with the sigmoid function that maps to the predicted CTR.
The decayed factor is set to $\lambda=0.5$.
We use the Adam optimizer for all methods. 
With grid search, the optimal learning rate is $1\times10^{-3}$ for AdaMoE and IADM, $1\times10^{-2}$ for IncCTR, and $3\times10^{-4}$ for MoE. 



\subsection{Quantitative Results}
\textbf{Industrial Dataset}
As shown in Table. \ref{tab:main-res}, AdaMoE outperforms all baseline models with \#exp. varying from three to twelve.
When the backbone is DIN and \#exp. is six, the overall AUC of AdaMoE improves by 0.19\% compared to the best baseline method (IADM with 3 experts).
The improvement of AUC is significant in CTR prediction \cite{wang2017deep,cheng2016wide,guo2017deepfm}.
We obtain the best result when \#exp. is six for AdaMoE.
However, for IADM we can see that too many experts can result in performance degradation. 
since IADM ensembles experts vertically which can be problematic when depth of the network grows larger.
The case of the single expert (\#exp.=1) is plotted in Fig. \ref{fig:auc-models} since all methods degenerate to the same. All MoE based methods outperforms single expert network as expected.

\textbf{Public Dataset}. 
The proposed AdaMoE also performs better than other baseline methods at any \#exp.. 
Since Avazu scales much smaller than industrial datasets, training based on gradient descent might be not efficient, especially for models with more parameters. While for AdaMoE, the update of expert weights is in closed-form and thereby more efficient. 
Besides, we can see that AUC generally drops as \#exp. increases for baselines.
One possible reason is that due to the small scale of the public dataset, as \#exp. increases, the underfitting degree the model gradually increases.

\subsection{Qualitative Results}

To further study how the models handle the concept drift in streaming data, we plot the AUC/10Min in Fig. \ref{fig:auc-models}.
The \#exp. is set to six and the backbone model is DIN.
Similar results can be observed with other settings and are omitted here.
We marked 18:40 in Fig. \ref{fig:auc-models}, which corresponds to the occurrence of concept drift.
Before 18:40, the AUC/10Min of AdaMoE increases faster than other baselines, which means AdaMoE can quickly learn with stream data.
After 18:40, the AUC/10Min of all methods begins to drop.
Among all models, AdaMoE first stops the drop, and its decline is about 30\%-40\% less than other baselines, which means AdaMoE quickly adapts to new distribution and alleviates concept drift.


\subsection{Influence of Decayed Factor $\lambda$.}
We investigate the impact of the decayed factor $\lambda$ in Fig. \ref{fig:auc-lambda}.
When $\lambda$ is small, AdaMoE can quickly adapt to stream data, but
AUC/10Min fluctuates more and the model is unstable.
We empirically find $\lambda=0.5$ achieves a better tradeoff between the stability and adaptability of AdaMoE, and yields the best overall AUC.

\subsection{Online A/B Testing}
The proposed AdaMoE model has been deployed in the online advertising system of one of the world's largest E-commerce companies.
The online A/B test lasts for a week (from 2022-Jan-7 to 2022-Jan-13).
Compared to the highly optimized base model, the proposed AdaMoE model contributes to 1.8\% CTR (Click Through rate) and 1.89\% eCPM (Effective Cost Per Mille) gain.





\section{Conclusion}
In this paper, we introduce a novel incremental learning framework, AdaMoE, to address the concept drift problem in CTR prediction.
The experiments show that our method overwhelms all other incremental learning methods considered both on a real-world production dataset and a reorganized public benchmark.
The online A/B test results further demonstrate the effectiveness of the proposed method.
Qualitative results and theoretical derivation are provided to illustrate the superiority of AdaMoE on drifting data streams. 

\balance
\bibliographystyle{ACM-Reference-Format}
\bibliography{sample-base}


\begin{thebibliography}{29}


\ifx \showCODEN    \undefined \def \showCODEN     #1{\unskip}     \fi
\ifx \showDOI      \undefined \def \showDOI       #1{#1}\fi
\ifx \showISBNx    \undefined \def \showISBNx     #1{\unskip}     \fi
\ifx \showISBNxiii \undefined \def \showISBNxiii  #1{\unskip}     \fi
\ifx \showISSN     \undefined \def \showISSN      #1{\unskip}     \fi
\ifx \showLCCN     \undefined \def \showLCCN      #1{\unskip}     \fi
\ifx \shownote     \undefined \def \shownote      #1{#1}          \fi
\ifx \showarticletitle \undefined \def \showarticletitle #1{#1}   \fi
\ifx \showURL      \undefined \def \showURL       {\relax}        \fi
\providecommand\bibfield[2]{#2}
\providecommand\bibinfo[2]{#2}
\providecommand\natexlab[1]{#1}
\providecommand\showeprint[2][]{arXiv:#2}

\bibitem[Bifet and Gavalda(2007)]%
        {bifet2007learning}
\bibfield{author}{\bibinfo{person}{Albert Bifet} {and} \bibinfo{person}{Ricard
  Gavalda}.} \bibinfo{year}{2007}\natexlab{}.
\newblock \showarticletitle{Learning from time-changing data with adaptive
  windowing}. In \bibinfo{booktitle}{\emph{Proceedings of the 2007 SIAM
  International Conference on Data Mining}}. \bibinfo{publisher}{{SIAM}},
  \bibinfo{address}{Minneapolis, Minnesota, {USA}}, \bibinfo{pages}{443--448}.
\newblock


\bibitem[Cheng et~al\mbox{.}(2016)]%
        {cheng2016wide}
\bibfield{author}{\bibinfo{person}{Heng-Tze Cheng}, \bibinfo{person}{Levent
  Koc}, \bibinfo{person}{Jeremiah Harmsen}, \bibinfo{person}{Tal Shaked},
  \bibinfo{person}{Tushar Chandra}, \bibinfo{person}{Hrishi Aradhye},
  \bibinfo{person}{Glen Anderson}, \bibinfo{person}{Greg Corrado},
  \bibinfo{person}{Wei Chai}, \bibinfo{person}{Mustafa Ispir}, {et~al\mbox{.}}}
  \bibinfo{year}{2016}\natexlab{}.
\newblock \showarticletitle{Wide \& deep learning for recommender systems}. In
  \bibinfo{booktitle}{\emph{Proceedings of the 1st Workshop on Deep Learning
  for Recommender Systems}}. \bibinfo{publisher}{{ACM}},
  \bibinfo{address}{Boston, MA, USA}, \bibinfo{pages}{7--10}.
\newblock


\bibitem[Eigen et~al\mbox{.}(2014)]%
        {eigen2013learning}
\bibfield{author}{\bibinfo{person}{David Eigen}, \bibinfo{person}{Marc'Aurelio
  Ranzato}, {and} \bibinfo{person}{Ilya Sutskever}.}
  \bibinfo{year}{2014}\natexlab{}.
\newblock \showarticletitle{Learning factored representations in a deep mixture
  of experts}. In \bibinfo{booktitle}{\emph{the 2nd International Conference on
  Learning Representations, Workshop Track Proceedings}}.
  \bibinfo{address}{Banff, AB, Canada}.
\newblock


\bibitem[Elwell and Polikar(2011)]%
        {elwell2011incremental}
\bibfield{author}{\bibinfo{person}{Ryan Elwell} {and} \bibinfo{person}{Robi
  Polikar}.} \bibinfo{year}{2011}\natexlab{}.
\newblock \showarticletitle{Incremental learning of concept drift in
  nonstationary environments}.
\newblock \bibinfo{journal}{\emph{{IEEE} Transactions on Neural Networks}}
  \bibinfo{volume}{22}, \bibinfo{number}{10} (\bibinfo{year}{2011}),
  \bibinfo{pages}{1517--1531}.
\newblock


\bibitem[Gepperth and Hammer(2016)]%
        {gepperth2016incremental}
\bibfield{author}{\bibinfo{person}{Alexander Gepperth} {and}
  \bibinfo{person}{Barbara Hammer}.} \bibinfo{year}{2016}\natexlab{}.
\newblock \showarticletitle{Incremental learning algorithms and applications}.
  In \bibinfo{booktitle}{\emph{European Symposium on Artificial Neural Networks
  (ESANN)}}. \bibinfo{address}{Bruges, Belgium}.
\newblock


\bibitem[Gomes et~al\mbox{.}(2017)]%
        {gomes2017survey}
\bibfield{author}{\bibinfo{person}{Heitor~Murilo Gomes},
  \bibinfo{person}{Jean~Paul Barddal}, \bibinfo{person}{Fabr{\'{\i}}cio
  Enembreck}, {and} \bibinfo{person}{Albert Bifet}.}
  \bibinfo{year}{2017}\natexlab{}.
\newblock \showarticletitle{A survey on ensemble learning for data stream
  Classification}.
\newblock \bibinfo{journal}{\emph{{ACM} Comput. Surv.}} \bibinfo{volume}{50},
  \bibinfo{number}{2} (\bibinfo{year}{2017}), \bibinfo{pages}{23:1--23:36}.
\newblock


\bibitem[Guo et~al\mbox{.}(2017)]%
        {guo2017deepfm}
\bibfield{author}{\bibinfo{person}{Huifeng Guo}, \bibinfo{person}{Ruiming
  Tang}, \bibinfo{person}{Yunming Ye}, \bibinfo{person}{Zhenguo Li}, {and}
  \bibinfo{person}{Xiuqiang He}.} \bibinfo{year}{2017}\natexlab{}.
\newblock \showarticletitle{DeepFM: {A} Factorization-Machine based Neural
  Network for {CTR} Prediction}. In \bibinfo{booktitle}{\emph{Proceedings of
  the 26th International Joint Conference on Artificial Intelligence}}.
  \bibinfo{publisher}{ijcai.org}, \bibinfo{address}{Melbourne, Australia},
  \bibinfo{pages}{1725--1731}.
\newblock


\bibitem[Huang et~al\mbox{.}(2021)]%
        {huang2021deep}
\bibfield{author}{\bibinfo{person}{Zai Huang}, \bibinfo{person}{Mingyuan Tao},
  {and} \bibinfo{person}{Bufeng Zhang}.} \bibinfo{year}{2021}\natexlab{}.
\newblock \showarticletitle{Deep User Match Network for Click-Through Rate
  Prediction}. In \bibinfo{booktitle}{\emph{proceedings of the 44th {ACM}
  {SIGIR} Conference on Research and Development in Information Retrieval}}.
  \bibinfo{publisher}{{ACM}}, \bibinfo{address}{Virtual Event, Canada},
  \bibinfo{pages}{1890--1894}.
\newblock


\bibitem[Jacobs et~al\mbox{.}(1991)]%
        {jacobs1991adaptive}
\bibfield{author}{\bibinfo{person}{Robert~A Jacobs}, \bibinfo{person}{Michael~I
  Jordan}, \bibinfo{person}{Steven~J Nowlan}, {and} \bibinfo{person}{Geoffrey~E
  Hinton}.} \bibinfo{year}{1991}\natexlab{}.
\newblock \showarticletitle{Adaptive mixtures of local experts}.
\newblock \bibinfo{journal}{\emph{Neural computation}} \bibinfo{volume}{3},
  \bibinfo{number}{1} (\bibinfo{year}{1991}), \bibinfo{pages}{79--87}.
\newblock


\bibitem[Kolter and Maloof(2007)]%
        {kolter2007dynamic}
\bibfield{author}{\bibinfo{person}{J.~Zico Kolter} {and}
  \bibinfo{person}{Marcus~A. Maloof}.} \bibinfo{year}{2007}\natexlab{}.
\newblock \showarticletitle{Dynamic weighted majority: {A}n ensemble method for
  drifting concepts}.
\newblock \bibinfo{journal}{\emph{Journal of Machine Learning Research}}
  \bibinfo{volume}{8} (\bibinfo{year}{2007}), \bibinfo{pages}{2755--2790}.
\newblock


\bibitem[Krawczyk et~al\mbox{.}(2017)]%
        {krawczyk2017ensemble}
\bibfield{author}{\bibinfo{person}{Bartosz Krawczyk},
  \bibinfo{person}{Leandro~L Minku}, \bibinfo{person}{Jo{\~a}o Gama},
  \bibinfo{person}{Jerzy Stefanowski}, {and} \bibinfo{person}{Micha{\l}
  Wo{\'z}niak}.} \bibinfo{year}{2017}\natexlab{}.
\newblock \showarticletitle{Ensemble learning for data stream analysis: A
  survey}.
\newblock \bibinfo{journal}{\emph{Information Fusion}}  \bibinfo{volume}{37}
  (\bibinfo{year}{2017}), \bibinfo{pages}{132--156}.
\newblock


\bibitem[Lu et~al\mbox{.}(2018)]%
        {lu2018learning}
\bibfield{author}{\bibinfo{person}{Jie Lu}, \bibinfo{person}{Anjin Liu},
  \bibinfo{person}{Fan Dong}, \bibinfo{person}{Feng Gu}, \bibinfo{person}{Joao
  Gama}, {and} \bibinfo{person}{Guangquan Zhang}.}
  \bibinfo{year}{2018}\natexlab{}.
\newblock \showarticletitle{Learning under concept drift: A review}.
\newblock \bibinfo{journal}{\emph{IEEE Transactions on Knowledge and Data
  Engineering}} \bibinfo{volume}{31}, \bibinfo{number}{12}
  (\bibinfo{year}{2018}), \bibinfo{pages}{2346--2363}.
\newblock


\bibitem[Lu et~al\mbox{.}(2017)]%
        {yang2017dynamic}
\bibfield{author}{\bibinfo{person}{Yang Lu}, \bibinfo{person}{Yiu{-}ming
  Cheung}, {and} \bibinfo{person}{Yuan~Yan Tang}.}
  \bibinfo{year}{2017}\natexlab{}.
\newblock \showarticletitle{Dynamic weighted majority for incremental learning
  of imbalanced data streams with concept drift}. In
  \bibinfo{booktitle}{\emph{Proceedings of the 26th International Joint
  Conference on Artificial Intelligence}}. \bibinfo{publisher}{ijcai.org},
  \bibinfo{address}{Melbourne, Australia}, \bibinfo{pages}{2393--2399}.
\newblock


\bibitem[Makkuva et~al\mbox{.}(2019)]%
        {makkuva2019breaking}
\bibfield{author}{\bibinfo{person}{Ashok Makkuva}, \bibinfo{person}{Pramod
  Viswanath}, \bibinfo{person}{Sreeram Kannan}, {and} \bibinfo{person}{Sewoong
  Oh}.} \bibinfo{year}{2019}\natexlab{}.
\newblock \showarticletitle{Breaking the gridlock in mixture-of-experts:
  Consistent and efficient algorithms}. In
  \bibinfo{booktitle}{\emph{International Conference on Machine Learning}}.
  PMLR, \bibinfo{pages}{4304--4313}.
\newblock


\bibitem[Qu et~al\mbox{.}(2016)]%
        {qu2016product}
\bibfield{author}{\bibinfo{person}{Yanru Qu}, \bibinfo{person}{Han Cai},
  \bibinfo{person}{Kan Ren}, \bibinfo{person}{Weinan Zhang},
  \bibinfo{person}{Yong Yu}, \bibinfo{person}{Ying Wen}, {and}
  \bibinfo{person}{Jun Wang}.} \bibinfo{year}{2016}\natexlab{}.
\newblock \showarticletitle{Product-based neural networks for user response
  prediction}. In \bibinfo{booktitle}{\emph{{IEEE} 16th International
  Conference on Data Mining ({ICDM})}}. \bibinfo{publisher}{{IEEE} Computer
  Society}, \bibinfo{address}{Barcelona, Spain}, \bibinfo{pages}{1149--1154}.
\newblock


\bibitem[Ramanath et~al\mbox{.}(2021)]%
        {ramanath2021lambda}
\bibfield{author}{\bibinfo{person}{Rohan Ramanath}, \bibinfo{person}{Konstantin
  Salomatin}, \bibinfo{person}{Jeffrey~D Gee}, \bibinfo{person}{Kirill
  Talanine}, \bibinfo{person}{Onkar Dalal}, \bibinfo{person}{Gungor Polatkan},
  \bibinfo{person}{Sara Smoot}, {and} \bibinfo{person}{Deepak Kumar}.}
  \bibinfo{year}{2021}\natexlab{}.
\newblock \showarticletitle{Lambda Learner: Fast Incremental Learning on Data
  Streams}. In \bibinfo{booktitle}{\emph{Proceedings of the 27th ACM SIGKDD
  Conference on Knowledge Discovery \& Data Mining}}.
  \bibinfo{pages}{3492--3502}.
\newblock


\bibitem[Ram{\'\i}rez-Gallego et~al\mbox{.}(2017)]%
        {ramirez2017survey}
\bibfield{author}{\bibinfo{person}{Sergio Ram{\'\i}rez-Gallego},
  \bibinfo{person}{Bartosz Krawczyk}, \bibinfo{person}{Salvador Garc{\'\i}a},
  \bibinfo{person}{Micha{\l} Wo{\'z}niak}, {and} \bibinfo{person}{Francisco
  Herrera}.} \bibinfo{year}{2017}\natexlab{}.
\newblock \showarticletitle{A survey on data preprocessing for data stream
  mining: Current status and future directions}.
\newblock \bibinfo{journal}{\emph{Neurocomputing}}  \bibinfo{volume}{239}
  (\bibinfo{year}{2017}), \bibinfo{pages}{39--57}.
\newblock


\bibitem[Shazeer et~al\mbox{.}(2017)]%
        {ShazeerMMDLHD17}
\bibfield{author}{\bibinfo{person}{Noam Shazeer}, \bibinfo{person}{Azalia
  Mirhoseini}, \bibinfo{person}{Krzysztof Maziarz}, \bibinfo{person}{Andy
  Davis}, \bibinfo{person}{Quoc~V. Le}, \bibinfo{person}{Geoffrey~E. Hinton},
  {and} \bibinfo{person}{Jeff Dean}.} \bibinfo{year}{2017}\natexlab{}.
\newblock \showarticletitle{Outrageously large neural networks: The
  sparsely-gated mixture-of-experts Layer}. In \bibinfo{booktitle}{\emph{the
  5th International Conference on Learning Representations, {ICLR}}}.
  \bibinfo{publisher}{OpenReview.net}, \bibinfo{address}{Toulon, France}.
\newblock


\bibitem[Street and Kim(2001)]%
        {street2001streaming}
\bibfield{author}{\bibinfo{person}{W.~Nick Street} {and}
  \bibinfo{person}{YongSeog Kim}.} \bibinfo{year}{2001}\natexlab{}.
\newblock \showarticletitle{A streaming ensemble algorithm {(SEA)} for
  large-scale classification}. In \bibinfo{booktitle}{\emph{Proceedings of the
  7th {ACM} {SIGKDD} international conference on Knowledge discovery \& data
  mining}}. \bibinfo{publisher}{{ACM}}, \bibinfo{address}{Francisco, CA, USA},
  \bibinfo{pages}{377--382}.
\newblock


\bibitem[Sugiyama(2015)]%
        {sugiyama2015introduction}
\bibfield{author}{\bibinfo{person}{Masashi Sugiyama}.}
  \bibinfo{year}{2015}\natexlab{}.
\newblock \bibinfo{booktitle}{\emph{Introduction to statistical machine
  learning}}.
\newblock \bibinfo{publisher}{Morgan Kaufmann}.
\newblock


\bibitem[Tsymbal(2004)]%
        {tsymbal2004problem}
\bibfield{author}{\bibinfo{person}{Alexey Tsymbal}.}
  \bibinfo{year}{2004}\natexlab{}.
\newblock \showarticletitle{The problem of concept drift: {D}efinitions and
  related work}.
\newblock \bibinfo{journal}{\emph{Computer Science Department, Trinity College
  Dublin}} \bibinfo{volume}{106}, \bibinfo{number}{2} (\bibinfo{year}{2004}),
  \bibinfo{pages}{58}.
\newblock


\bibitem[Wang et~al\mbox{.}(2017)]%
        {wang2017deep}
\bibfield{author}{\bibinfo{person}{Ruoxi Wang}, \bibinfo{person}{Bin Fu},
  \bibinfo{person}{Gang Fu}, {and} \bibinfo{person}{Mingliang Wang}.}
  \bibinfo{year}{2017}\natexlab{}.
\newblock \showarticletitle{Deep {\&} cross network for ad click predictions}.
  In \bibinfo{booktitle}{\emph{Proceedings of the ADKDD}}.
  \bibinfo{publisher}{{ACM}}, \bibinfo{address}{Halifax, NS, Canada},
  \bibinfo{pages}{12:1--12:7}.
\newblock


\bibitem[Wang et~al\mbox{.}(2020)]%
        {wang2020practical}
\bibfield{author}{\bibinfo{person}{Yichao Wang}, \bibinfo{person}{Huifeng Guo},
  \bibinfo{person}{Ruiming Tang}, \bibinfo{person}{Zhirong Liu}, {and}
  \bibinfo{person}{Xiuqiang He}.} \bibinfo{year}{2020}\natexlab{}.
\newblock \showarticletitle{A practical incremental method to train deep ctr
  models}.
\newblock \bibinfo{journal}{\emph{arXiv preprint arXiv:2009.02147}}
  (\bibinfo{year}{2020}).
\newblock


\bibitem[Widmer and Kubat(1996)]%
        {Widmer1996learning}
\bibfield{author}{\bibinfo{person}{Gerhard Widmer} {and}
  \bibinfo{person}{Miroslav Kubat}.} \bibinfo{year}{1996}\natexlab{}.
\newblock \showarticletitle{Learning in the Presence of Concept Drift and
  Hidden Contexts}.
\newblock \bibinfo{journal}{\emph{Machine Learning}} \bibinfo{volume}{23},
  \bibinfo{number}{1} (\bibinfo{year}{1996}), \bibinfo{pages}{69--101}.
\newblock


\bibitem[Xiao et~al\mbox{.}(2020)]%
        {xiao2020deep}
\bibfield{author}{\bibinfo{person}{Zhibo Xiao}, \bibinfo{person}{Luwei Yang},
  \bibinfo{person}{Wen Jiang}, \bibinfo{person}{Yi Wei}, \bibinfo{person}{Yi
  Hu}, {and} \bibinfo{person}{Hao Wang}.} \bibinfo{year}{2020}\natexlab{}.
\newblock \showarticletitle{Deep multi-interest network for click-through rate
  prediction}. In \bibinfo{booktitle}{\emph{Proceedings of the 29th {ACM}
  International Conference on Information and Knowledge Management (CIKM)}}.
  \bibinfo{publisher}{{ACM}}, \bibinfo{address}{Virtual Event, Ireland},
  \bibinfo{pages}{2265--2268}.
\newblock


\bibitem[Yang et~al\mbox{.}(2019)]%
        {yang2019adaptive}
\bibfield{author}{\bibinfo{person}{Yang Yang}, \bibinfo{person}{Da{-}Wei Zhou},
  \bibinfo{person}{De{-}Chuan Zhan}, \bibinfo{person}{Hui Xiong}, {and}
  \bibinfo{person}{Yuan Jiang}.} \bibinfo{year}{2019}\natexlab{}.
\newblock \showarticletitle{Adaptive deep models for incremental learning:
  {C}onsidering capacity scalability and sustainability}. In
  \bibinfo{booktitle}{\emph{Proceedings of the 25th {ACM} {SIGKDD}
  International Conference on Knowledge Discovery {\&} Data Mining ({KDD})}}.
  \bibinfo{publisher}{{ACM}}, \bibinfo{address}{Anchorage, AK, USA},
  \bibinfo{pages}{74--82}.
\newblock


\bibitem[Zeevi et~al\mbox{.}(1998)]%
        {zeevi1998moe}
\bibfield{author}{\bibinfo{person}{Assaf Zeevi}, \bibinfo{person}{Ron Meir},
  {and} \bibinfo{person}{Vitaly Maiorov}.} \bibinfo{year}{1998}\natexlab{}.
\newblock \showarticletitle{Error Bounds for Functional Approximationand
  Estimation Using Mixtures of Experts}.
\newblock \bibinfo{journal}{\emph{IEEE Transactions on Information Theory}}
  \bibinfo{volume}{44}, \bibinfo{number}{3} (\bibinfo{year}{1998}),
  \bibinfo{pages}{1010--1025}.
\newblock


\bibitem[Zhou et~al\mbox{.}(2019)]%
        {zhou2019deep}
\bibfield{author}{\bibinfo{person}{Guorui Zhou}, \bibinfo{person}{Na Mou},
  \bibinfo{person}{Ying Fan}, \bibinfo{person}{Qi Pi}, \bibinfo{person}{Weijie
  Bian}, \bibinfo{person}{Chang Zhou}, \bibinfo{person}{Xiaoqiang Zhu}, {and}
  \bibinfo{person}{Kun Gai}.} \bibinfo{year}{2019}\natexlab{}.
\newblock \showarticletitle{Deep interest evolution network for click-through
  rate prediction}. In \bibinfo{booktitle}{\emph{The 33rd {AAAI} Conference on
  Artificial Intelligence ({AAAI})}}. \bibinfo{publisher}{{AAAI} Press},
  \bibinfo{address}{Honolulu, Hawaii, USA}, \bibinfo{pages}{5941--5948}.
\newblock


\bibitem[Zhou et~al\mbox{.}(2018)]%
        {zhou2018deep}
\bibfield{author}{\bibinfo{person}{Guorui Zhou}, \bibinfo{person}{Xiaoqiang
  Zhu}, \bibinfo{person}{Chengru Song}, \bibinfo{person}{Ying Fan},
  \bibinfo{person}{Han Zhu}, \bibinfo{person}{Xiao Ma},
  \bibinfo{person}{Yanghui Yan}, \bibinfo{person}{Junqi Jin},
  \bibinfo{person}{Han Li}, {and} \bibinfo{person}{Kun Gai}.}
  \bibinfo{year}{2018}\natexlab{}.
\newblock \showarticletitle{Deep interest network for click-through rate
  prediction}. In \bibinfo{booktitle}{\emph{Proceedings of the 24th {ACM}
  {SIGKDD} International Conference on Knowledge Discovery {\&} Data Mining
  ({KDD})}}. \bibinfo{publisher}{{ACM}}, \bibinfo{address}{London, UK},
  \bibinfo{pages}{1059--1068}.
\newblock


\end{thebibliography}










\end{document}